%% file: article.tex
\documentclass[a4paper,11pt]{article}
\pdfoutput=1

\usepackage{jinstpub}

\usepackage{tikz}
\usetikzlibrary{shapes.geometric}
\usetikzlibrary{arrows.meta}
\usepackage{lscape}
\usepackage{tikz-timing}

\usepackage{xcolor,colortbl}

\usepackage[utf8]{inputenc}

\usepackage{graphicx, color}

\usepackage{lipsum}

\title{Firmware implementation of a recurrent neural network for the computation of the energy deposited in the liquid argon calorimeter of the ATLAS experiment}

\author[a,1]{G. Aad,\note{Corresponding author.}}
\author[a]{T. Calvet,}
\author[a]{N. Chiedde,}
\author[a]{R. Faure,}
\author[a]{E.M. Fortin}
\author[b]{L. Laatu,}
\author[a]{E. Monnier,}
\author[a]{N. Sur}

\affiliation[a]{Aix Marseille Univ, CNRS/IN2P3, CPPM, Marseille, France}
\affiliation[b]{Aix Marseille Univ, CNRS/IN2P3, CPPM, IPhU, Marseille, France}

\emailAdd{aad@cern.ch}

\abstract{The ATLAS experiment measures the properties of particles that are products of proton-proton collisions at the LHC. The ATLAS detector will undergo a major upgrade before the high luminosity phase of the LHC. The ATLAS liquid argon calorimeter measures the energy of particles interacting electromagnetically in the detector. The readout electronics of this calorimeter will be replaced during the aforementioned ATLAS upgrade. The new electronic boards will be based on state-of-the-art field-programmable gate arrays (FPGA) from Intel allowing the implementation of neural networks embedded in firmware. Neural networks have been shown to outperform the current optimal filtering algorithms used to compute the energy deposited in the calorimeter. This article presents the implementation of a recurrent neural network (RNN) allowing the reconstruction of the energy deposited in the calorimeter on Stratix 10 FPGAs. The implementation in high level synthesis (HLS) language allowed fast prototyping but fell short of meeting the stringent requirements in terms of resource usage and latency. Further optimisations in Very High-Speed Integrated Circuit Hardware Description Language (VHDL) allowed fulfilment of the requirements of processing 384 channels per FPGA with a latency smaller than 125 ns. }

\keywords{Calorimeter methods, Digital signal processing (DSP), Data processing methods.}

\begin{document}
\maketitle
\flushbottom
	
\section{Introduction}
    
\begin{figure}[!htb]
    \centering
    \includegraphics[width=0.49\textwidth]{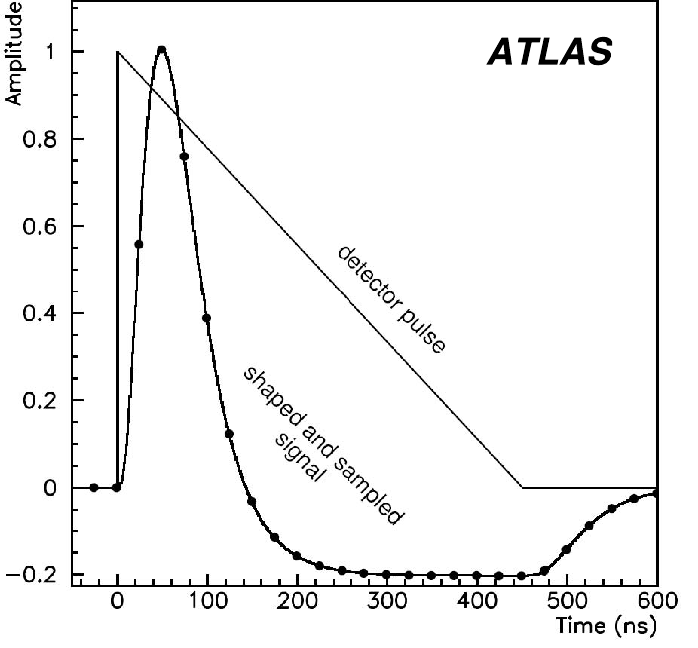}
    \caption{Shapes of the LAr calorimeter current pulse in the detector and of the signal output after bipolar shaping. The dots represent the samples seperated by 25 ns \cite{LAr-Phase-II-TDR}.}
    \label{fig:pulse_shape}
\end{figure}

The ATLAS experiment~\cite{atlas} at the Large Hadron Collider (LHC) \cite{LHC} measures the properties of particles produced in proton-proton collisions at energies of several teraelectronvolts [TeV] and a collision frequency of 40 MHz.
In the years 2026-2029 the LHC will undergo a major upgrade to increase its instantaneous luminosity by a factor of 5-7 leading to the High luminosity LHC (HL-LHC).
During the same period the ATLAS detector will be upgraded to cope with the increased luminosity of the HL-LHC.
This upgrade is called the phase-II upgrade.
The readout electronics of the ATLAS liquid argon (LAr) calorimeter will be replaced as part of the phase-II upgrade~\cite{LAr-Phase-II-TDR}.
The new frontend boards will shape, sample, and digitise at 40 MHz the electronic signal from the calorimeter before sending the samples to the backend electronics through optical fibers. Figure \ref{fig:pulse_shape} shows a typical pulse shape in the detector before and after the bipolar shaping. The new backend boards employ FPGAs to compute the energy deposited in the calorimeter out of the samples received from the frontend boards. The computed energy is then sent to the trigger system at 40 MHz, and to the readout system at 1 MHz in case of a level 1 trigger accept decision.
       
\begin{figure}[!htb]
    \centering
    \includegraphics[width=0.9\textwidth]{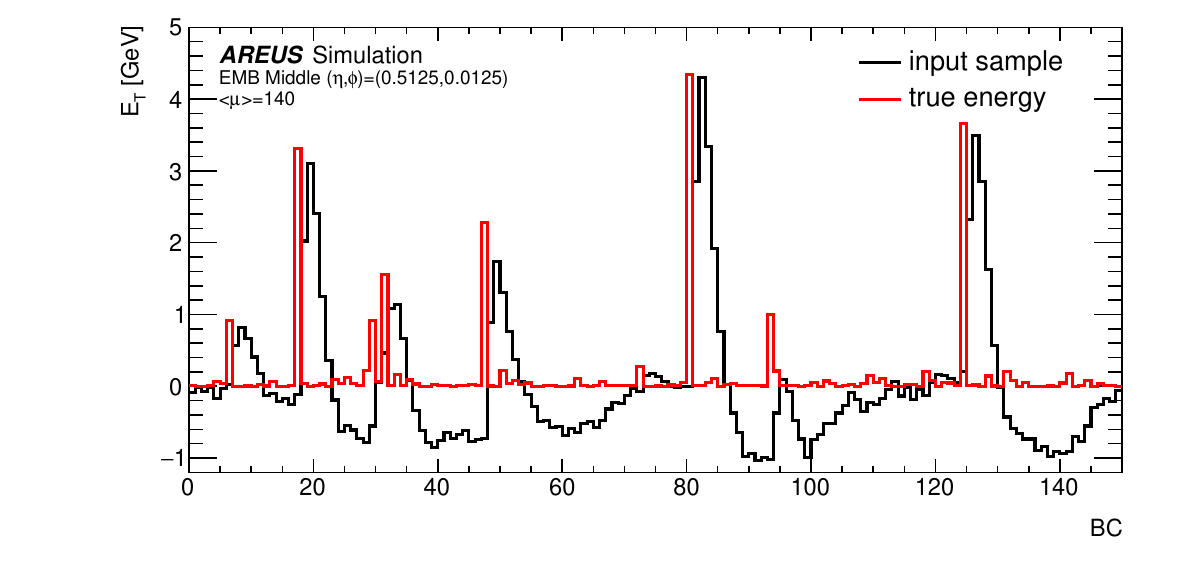}
    \caption{Sample sequence (black) of an EMB middle-layer cell located at a pseudorapidity $\eta$=0.5125 and an azimuthal angle $\phi$=0.0125 within the ATLAS coordinate system, simulated by AREUS, together with the true transverse energy (E$_\textrm{T}$) deposits (red), at an average pileup $\mu$ of 140 as a function of the bunch crossing (BC) counter. The samples amplitude is normalized to the value of the deposited energy in GeV.}
    \label{fig:pulse_train}
\end{figure}

The LAr calorimeter measures the energy of particles produced in LHC collisions. An excellent energy resolution and accurate detection of the energy-deposit time are crucial to enhance the ATLAS physics discovery potential at the HL-LHC. Currently, the transverse energy is computed using optimal filtering algorithms~\cite{of} that assume a nominal pulse shape of the electronic signal. Calorimeter electronic signals of up to 25 subsequent collisions overlap and create distortions to the pulse shape. This increases the difficulty of energy reconstruction and identification of the corresponding proton-proton bunch crossing. Up to 200 simultaneous proton-proton collisions are expected at the HL-LHC, which will lead to a high rate of overlapping signals in a given calorimeter channel. This will result in a significant energy degradation especially for low time-gap between two consecutive pulses, as discussed in~\cite{larnn}. Figure \ref{fig:pulse_train} shows a sequence of energy deposits in the calorimeter and the corresponding pulses simulated using AREUS \cite{areus}. The energy deposits correspond to events with a pileup of 140 collisions per bunch crossing. High energy deposits from hard scatter events are simulated by adding flat random energy deposits between zero and five GeV, separated by 30 bunch crossings on average.

Neural networks implemented in FPGAs have demonstrated enhanced object reconstruction and identification at trigger level in LHC experiments \cite{hls4ml,atlas_muon_fpga,atlas_muon_fpga2,cms_muon,nn_trigger}. The use of neural networks improves the energy resolution in the ATLAS LAr calorimeter especially in the low time-gap region as shown in~\cite{larnn}. Both convolutional neural networks (CNNs) and recurrent neural networks (RNNs) are shown to outperform the optimal filtering algorithm.
The transverse energy reconstruction is performed in custom electronic boards based on the latest state-of-the-art FPGAs~\cite{LAr-Phase-II-TDR}. Each FPGA should reconstruct the energies of 384 independent channels. The computation is done on-the-fly at the collision frequency of 40 MHz. Furthermore, the energy reconstruction should be done within a latency of 125 ns. These requirements put very stringent constraints on the firmware implementation of the energy reconstruction algorithms. The implementation should balance serialisation of the channels to save FPGA computation resources, and parallelisation to reduce the implementation latency and keep up with the high bandwidth. 

The electronic boards for the LAr calorimeter phase-II upgrade are currently under development and will use the Agilex~\cite{agilex} family of Intel FPGAs. A demonstrator board is already produced with Stratix 10 FPGAs~\cite{stratix}.
In this article we consider the implementation of a vanilla RNN~\cite{vanillarnn} algorithm in a Stratix 10 FPGA from Intel (part number 1SG280HU1F50E2VG). The implementation is carried out initially in HLS for fast prototyping and optimisation of the network parameters and the implementation architecture. After this initial implementation, VHDL is used for the final optimisation, ensuring that the requirements in terms of FPGA resource occupancy and latency are met.

\section{Vanilla RNN}

The neural network used in this article is the vanilla RNN with the sliding-window approach described in~\cite{larnn}. It is trained using Keras \cite{keras} with input simulated data from AREUS. At each bunch crossing, the network receives as input five consecutive samples of the electronic pulse of one channel of the LAr calorimeter and computes the corresponding deposited transverse energy. The five samples correspond to a time window of five bunch crossings. The computed transverse energy corresponds to a possible deposit at the second bunch crossing. Four of the samples are around the pulse peak generated by the deposited energy while the first sample is prior to the pulse allowing to detect the presence of overlapping pulses from past energy deposits.

\begin{figure}[!htb]
    \centering
    \resizebox{0.8\textwidth}{!}{
        \input{fig3}
    }
    \caption{Schematics of a Vanilla RNN with 5 cells and 8 internal dimensions (size of the state vector) followed by a dense layer. Each cell calculates a new state using the input data and the state of the previous cell. As the first cell does not have a previous cell, the calculation is only based on the input data. The dense layer computes the transverse energy using the state of the last cell.}
    \label{fig:Full_NN}
\end{figure}
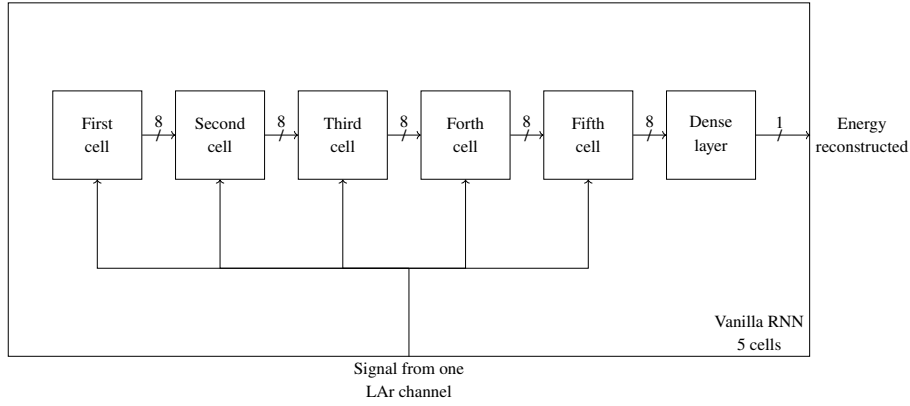

\begin{figure}[!htb]
    \centering
    \resizebox{1\textwidth}{!}{
        \includegraphics[width=0.49\textwidth]{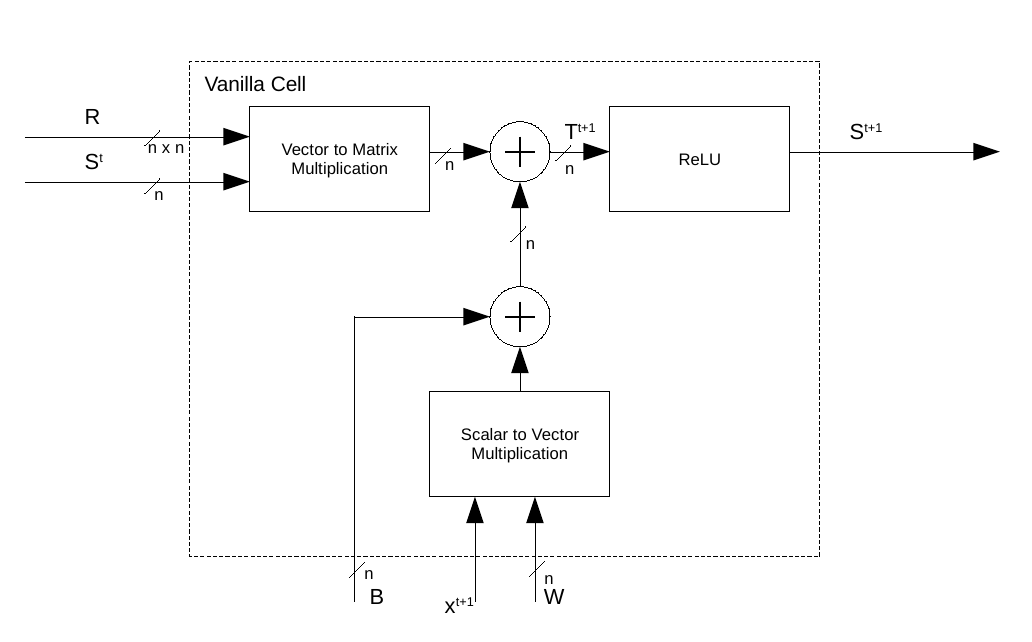}
    }
    \caption{Schematics of the operations performed inside each of the RNN cells. The recurrent weight multiplication multiplies the state vector from the previous cell (S$^\mathrm{t}$) with the recurrent kernel weight matrix (R). Simultaneously, the LAr cell input (X$^\mathrm{t+1}$) is multiplied by the kernel weight vector (W) and added to the bias weight (B). The results from the two above operations are added to create the internal vector T$^\mathrm{t+1}$. The ReLU activation function is applied on the elements of T$^\mathrm{t+1}$ to create the state vector S$^\mathrm{t+1}$. The state vector size (n) is equal to 8.}
    \label{fig:RNN}
\end{figure}

The network implementation is composed of five consecutive RNN cells followed by a dense layer as shown in figure~\ref{fig:Full_NN}. Each cell process the input from one sample corresponding to one bunch crossing. Each of the cells is composed of 6 blocks of computation as shown in figure~\ref{fig:RNN}: two addition blocks, one multiplication of a scalar with a vector, one multiplication of a vector with a matrix, and one activation function.
The weights of the network obtained from the training are stored in memory and are given as input to the RNN cells along with the electronic pulse samples. The same weights are used for each of the five network cells.
The ReLU, described in equation~\ref{eq:relu}, is used as the activation function for its simple implementation in FPGAs.
\begin{equation}
  f(x) =
    \begin{cases}
      0 & \text{if $x \leq 0$}\\
      x & \text{otherwise}
    \end{cases}  
    \label{eq:relu}
\end{equation}
    
\section{High-Level Synthesis Implementation}

The High Level Synthesis (HLS) is a design process that takes as input a behavioural specification of a digital system and convert it to a register-transfer level (RTL) realising the described functions. For the study presented in this article, the Intel HLS~\cite{hls} is used. The Intel HLS takes a C++ code as input and produces RTL optimised for Intel FPGAs.
It provides several macros which are inlined in the C++ code that allow control over the RTL implementation. The Intel HLS implements the standard C++ types but also defines several other types, in particular arbitrary precision fixed-point representations. 

Neural network cell computations are made of additions and multiplications. The aim of the firmware implementation is to reproduce with high precision the software computation results while keeping low resource usage in the FPGA and low latency.

\subsection{Implementation of multiplications}
\label{sec:multiplication}

Inside the Vanilla RNN cell there is one vector multiplication and one matrix multiplication which reduces to several vector multiplications which in turn reduce to several scalar multiplications and scalar additions.
There is a dedicated component inside the FPGA to perform scalar multiplications called the Digital Signal Processing (DSP). The DSP can be used in three possible modes.
The first mode performs one 32$\times$32 bits multiplication in the floating-point representation.
The second mode performs one 27$\times$27 bits multiplication in the fixed-point representation.
The third mode can perform two independent 19$\times$18 bits multiplications in the fixed-point representation.
As explained in~\ref{sec:multiplexing}, the number of multiplications of the FPGA limits the number of calorimeter cells that can be handled by one FPGA. The third mode is thus chosen since it allows doubling the available dedicated multiplication resources on the FPGA.

\subsection{Multiplexing}
\label{sec:multiplexing}
The number of multiplications inside the neural network depends on the size of the state vector ($n$) and the number of network cells ($c$) following equation~\ref{eq:nmulti} :
\begin{equation}
    n^2\times(c-1) + n\times(c+1)
    \label{eq:nmulti}
\end{equation}
where $n=8$ and $c=5$. Therefore, 304 multiplications are needed for the vanilla RNN used in this article. This leads to 116,736 multiplications that are needed to handle the 384 channels processed in one FPGA. The Stratix 10 FPGA possesses only 11,520 multiplication blocks. Therefore, each neural network instance in the FPGA must be utilised for several channels. The easiest way to do so is to increase the DSP block computation frequency as a multiple of the input frequency (40MHz). Several channels are then multiplexed using the same block. This multiplexing is limited by the maximum frequency of the computation block. In what follows we will target the maximum achievable frequency to increase the multiplexing for a given implementation of the neural network.

\subsection{Implementation of arithmetic operations}

Intel HLS implements two binary representations of numbers: a fixed point representation and a floating point representation. The implementation of floating point representation is relatively complex and uses more arithmetic and logic blocks for the addition operations. Furthermore, the DSP blocks of the Stratix 10 FPGA allows only one floating point multiplication instead of two simultaneous fixed point multiplications as explained in section~\ref{sec:multiplication}. The fixed point representation is chosen in order to minimise the resource usage in the FPGA.

The fixed point representation implemented in Intel HLS follows the Algorithmic C (AC) Data-types defined by Mentor Graphics~\cite{quantization}. Four parameters define this representation: the total number of bits, the number of bits of the integer part, the quantisation type, and the treatment of the overflow. Three different data categories are defined for the vanilla RNN implementation: the input and output data, the neural network weights, and the intermediate data which are the results of the internal computation inside the neural network blocks. 19 bits are used for the width of the internal and input/output categories, while 16 bits are used for the weights. This ensures an efficient use of the DSP resources inside the FPGA while providing a very good compatibility between the firmware computation and the one performed in the Keras software with floating points arithmetic operations. The firmware resolution, which is defined as the relative difference between the firmware and the Keras computed energies, is less than 0.1 \% with this choice of number of bits.

The overflow defines how the bits to the left of the most significant bit (MSB) are lost due to saturation. The number of bits and the position of the radix point are chosen to be able to represent the maximum value that could occur in the network. Thus, no saturation detection is needed and a simple drop of bits implementation is used.

 The quantisation defines how the bits to the right of the least significant bit (LSB) are lost. The loss of bits can occur during the conversion of the floating points representing the inputs and weights. The inputs to the neural network are given by the AREUS simulation while the weights are provided by Keras. These two use 32-bit signed floating point representations that need to be converted to a fixed point representation in order to be used in the firmware. The loss of bits can also occur inside the neural network internal computations. The 37-bit output of the DSP is reduced to the number of bits internally used. Two types of quantisation are implemented in the Intel HLS compilation: truncation and rounding. Each of these types possess different subtypes which are explained in~\cite{quantization}.
 
\begin{figure}[!htb]
    \centering
    \includegraphics[width=0.49\textwidth]{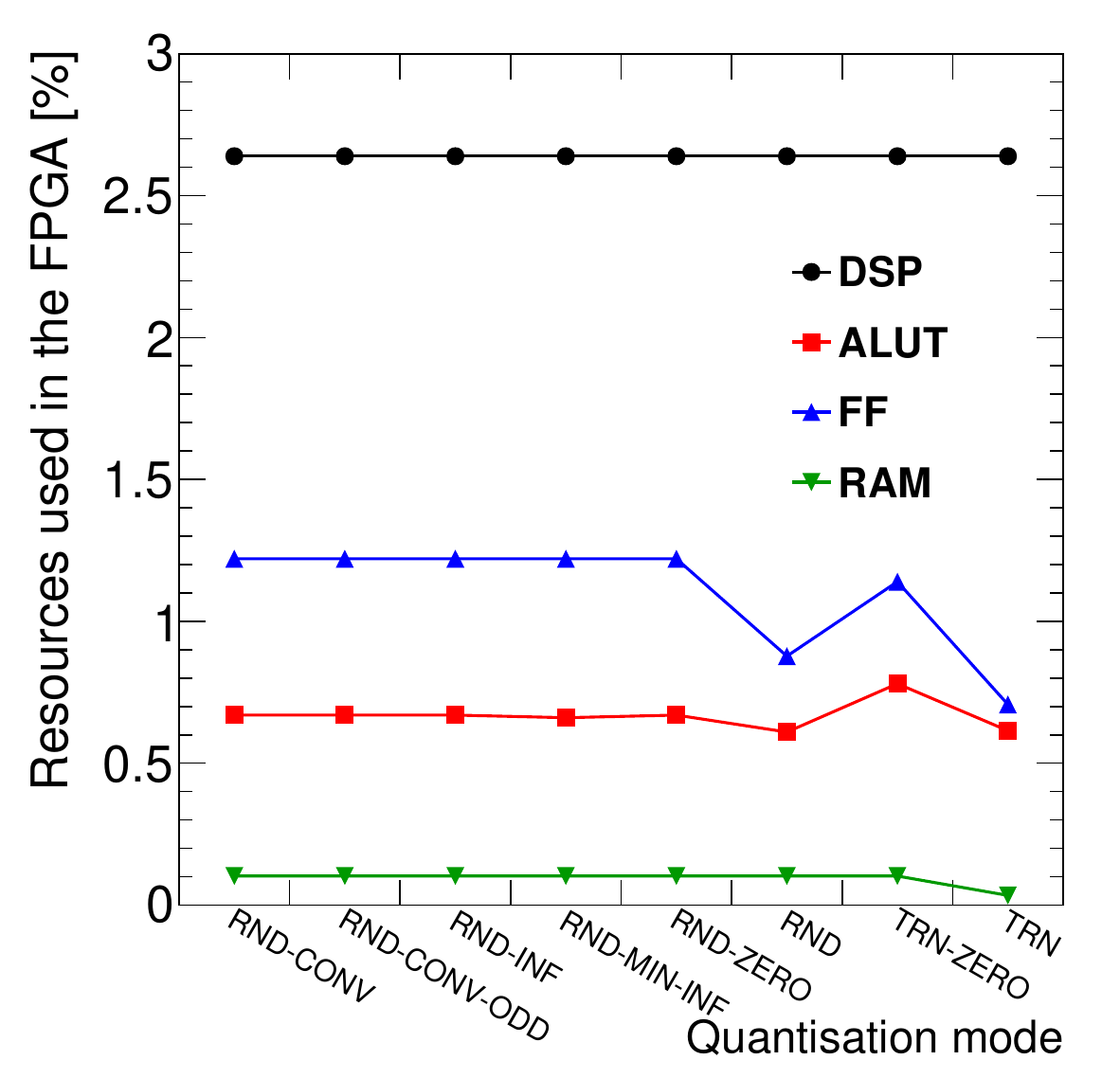}
    \includegraphics[width=0.49\textwidth]{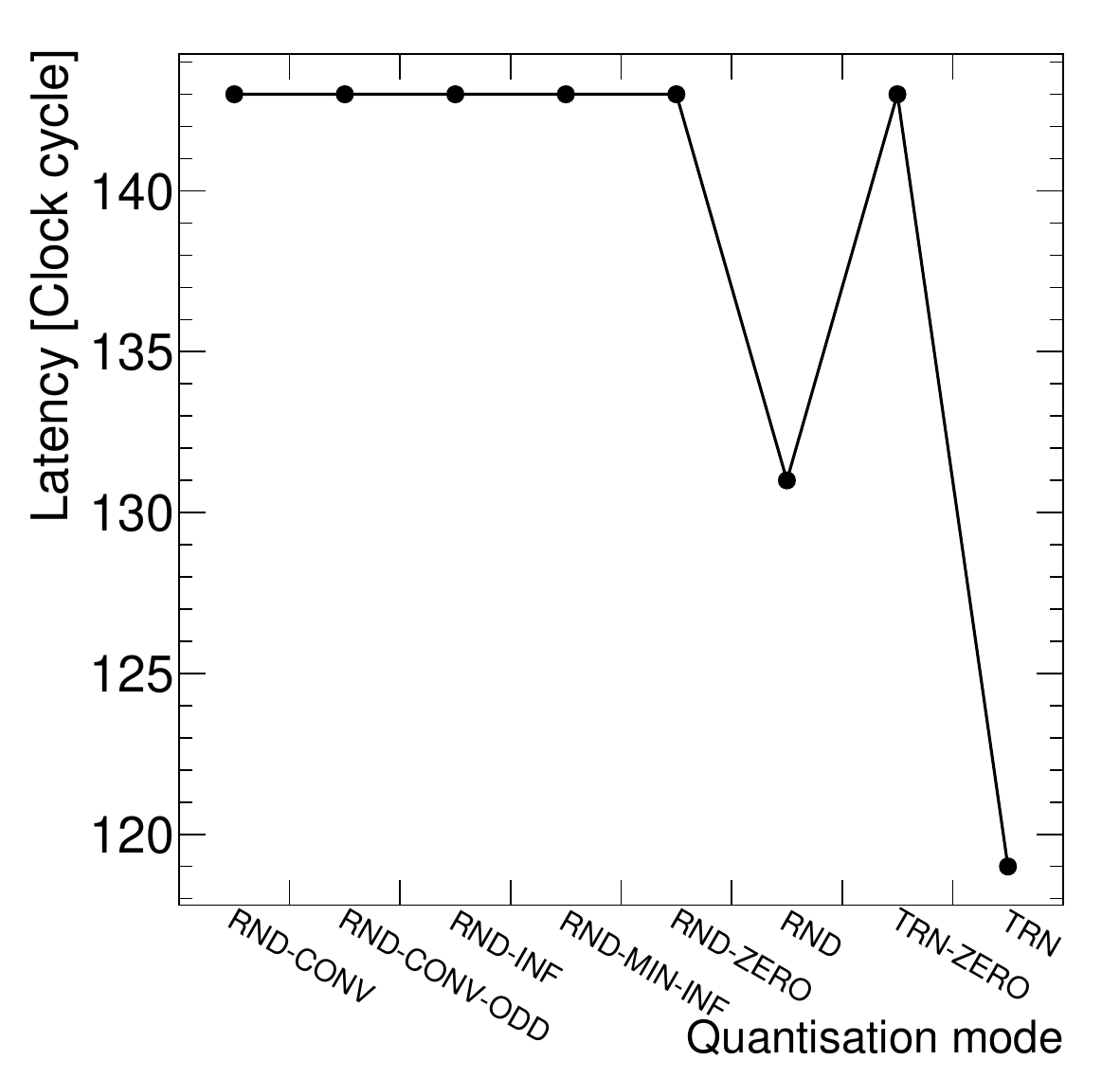}
    \caption{Comparison of the resource usage (left) and the latency (right) of the vanilla RNN firmware with different implementation of the quantisation in Intel HLS. For each mode, the same quantisation is applied for all categories of the data of the network. The different quantisation modes are described in~\cite{quantization}.}
    \label{fig:quantization}
\end{figure}

\begin{figure}[!htb]
    \centering
    \includegraphics[width=0.49\textwidth]{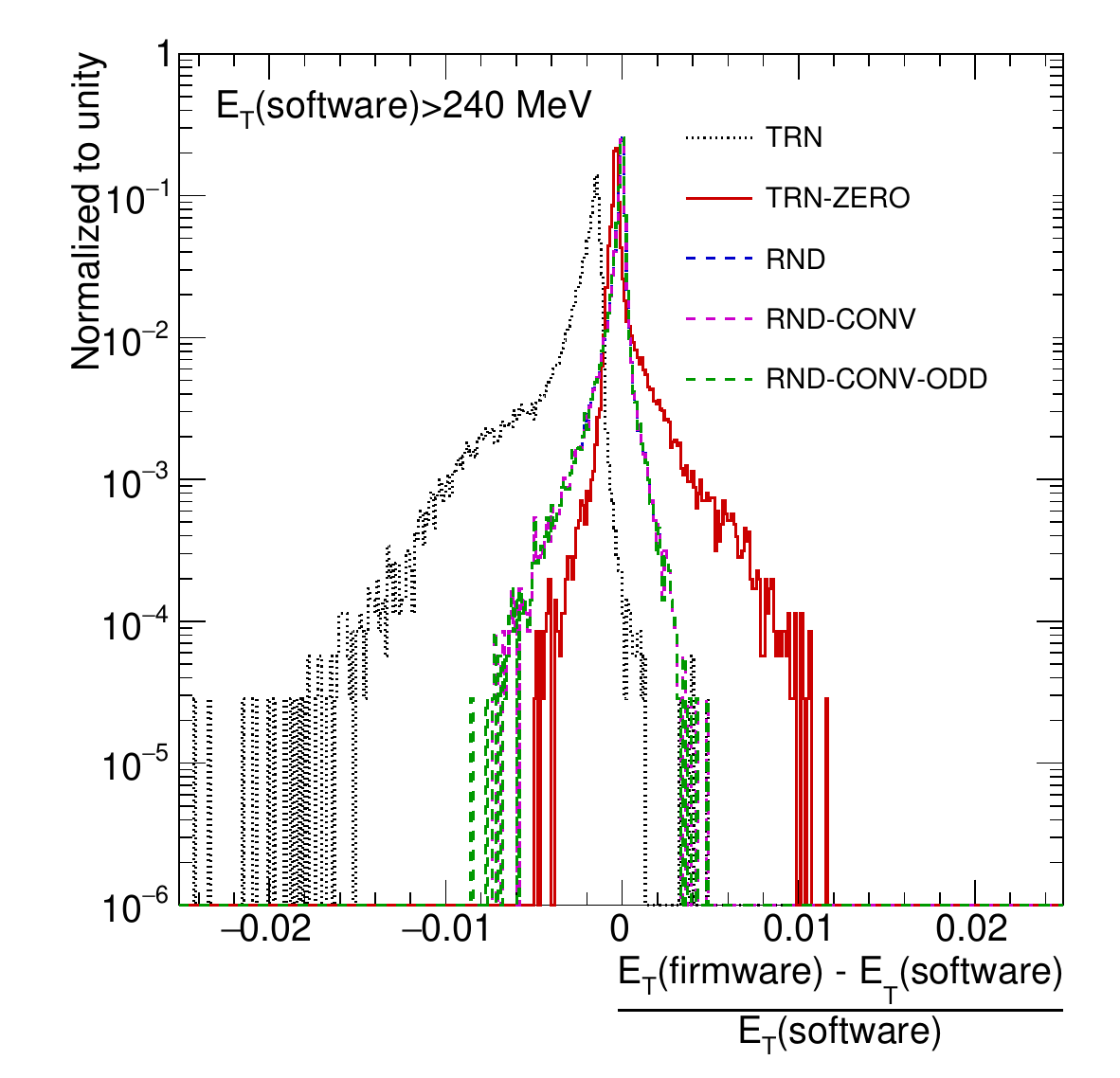}
    \includegraphics[width=0.49\textwidth]{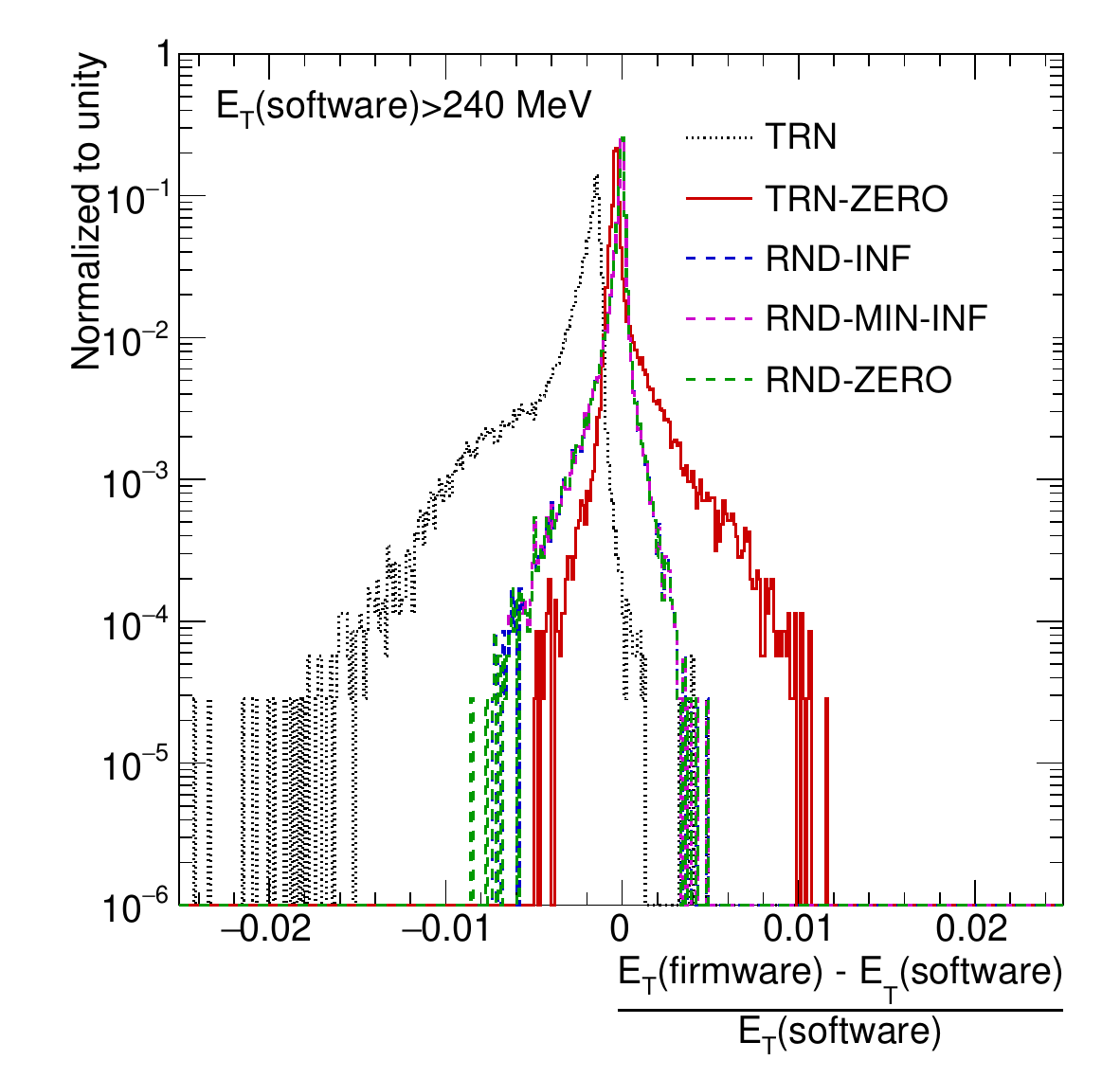}
    \caption{Resolution of the transverse energy (E$_\mathrm{T}$) computed in firmware with respect to the one computed in software. The E$_\mathrm{T}$(firmware) is computed with different implementations of the quantisation in Intel HLS.
    In each mode, the same quantisation is applied for all categories of the data of the network. The different quantisation modes are described in~\cite{quantization}. The different RND modes give very similar results and their corresponding curves overlap. A lower cut of 240 MeV is applied on $E_\mathrm{T}$(software) to remove low energies below the 3$\sigma$ noise level as described in~\cite{larnn}.}
    \label{fig:quantization_reso}
\end{figure}

Figure~\ref{fig:quantization} shows a comparison of the resource usage, in terms of DSPs, arithmetic lookup tables (ALUT), Flip Flops (FF), and random access memory (RAM), and the latency of the firmware for different implementations of the quantisation. The same quantisation is applied for all categories of data of the network. Two modes are interesting, the default truncation (TRN) and the default rounding (RND) which use lower resources and lower latency compared to other quantisation modes. Figure~\ref{fig:quantization_reso} shows a comparison of the transverse energy computed in firmware and the one computed in software, with a full floating point implementation, for the different quantisation modes. All quantisation modes show similar resolution with the exception of the TRN quantisation modes which have large tails. The RND mode gives a good compromise among resource usage, latency, and resolution. 
    

\begin{figure}[!htb]
    \centering
    \includegraphics[width=0.49\textwidth]{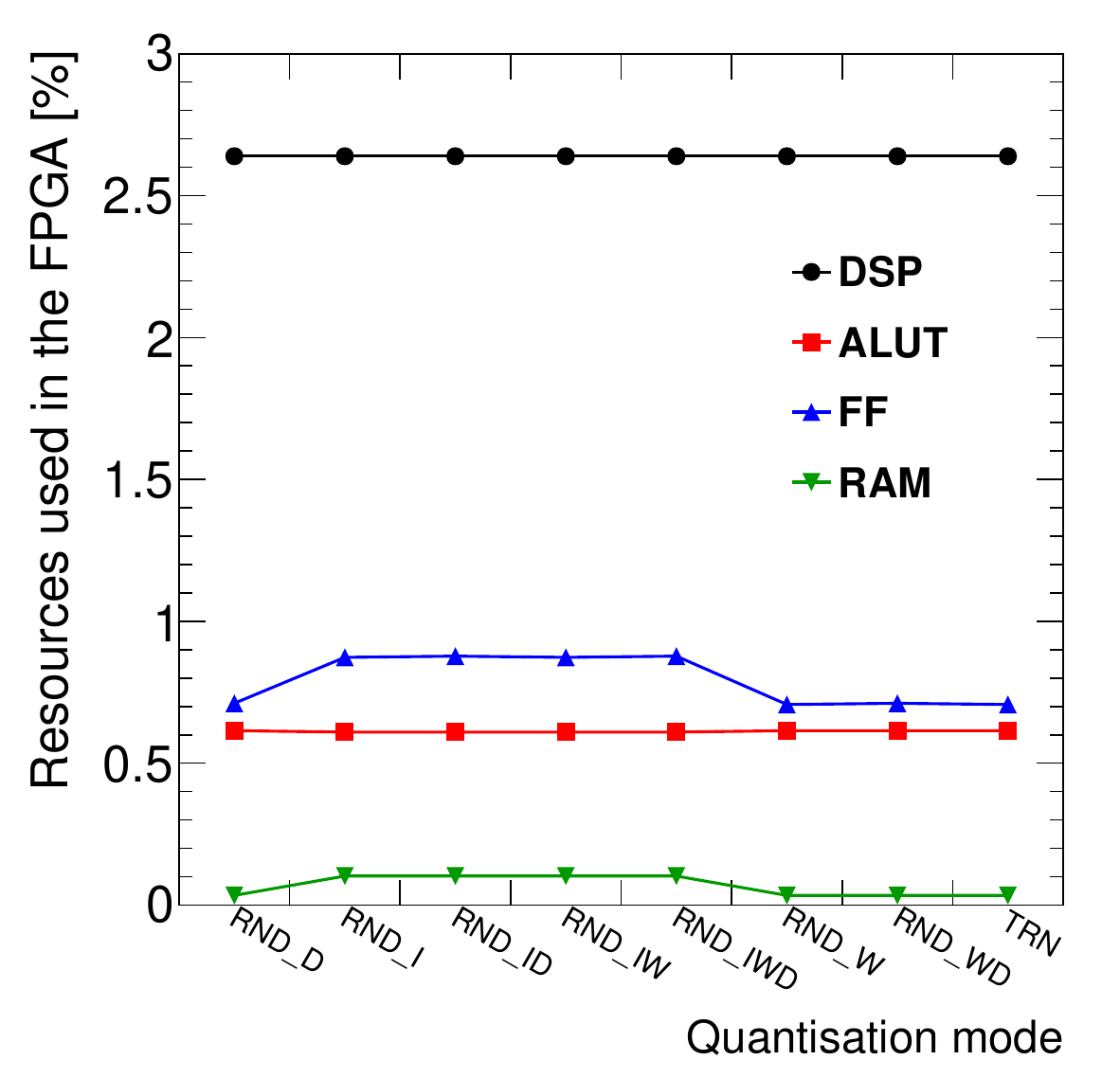}
    \includegraphics[width=0.49\textwidth]{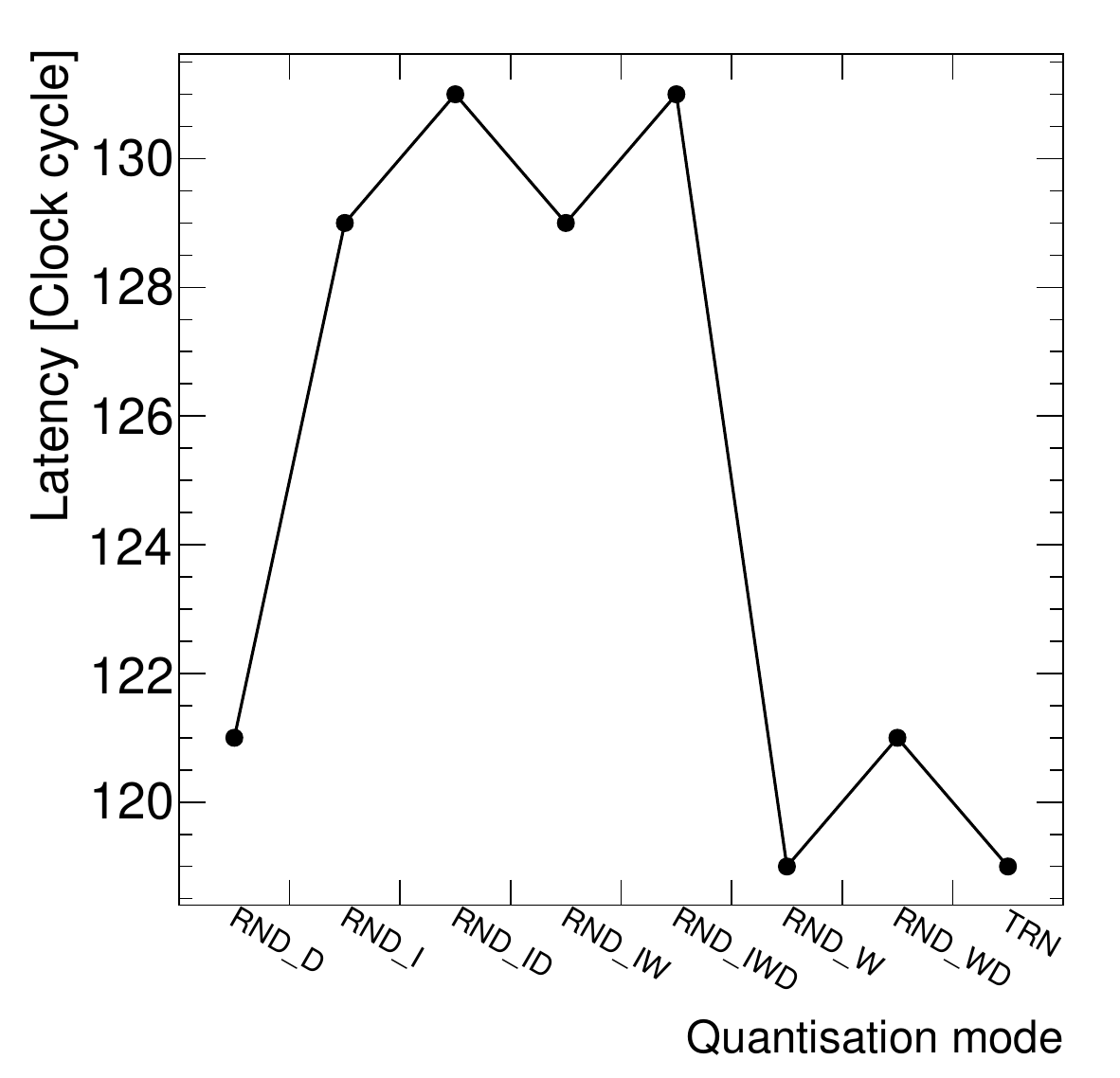}
    \caption{Comparison of the resource usage (left) and the latency (right) of the vanilla RNN firmware with different implementation of the quantisation for the three data categories of the network computation. For each test, the letters I, W, and D indicate that the RND is applied for the internal data category, the weights, and the input/output data, respectively, while the TRN mode is applied by default in all other categories.}
    \label{fig:quantization_diff_types}
\end{figure}

\begin{figure}[!htb]
    \centering
    \includegraphics[width=0.49\textwidth]{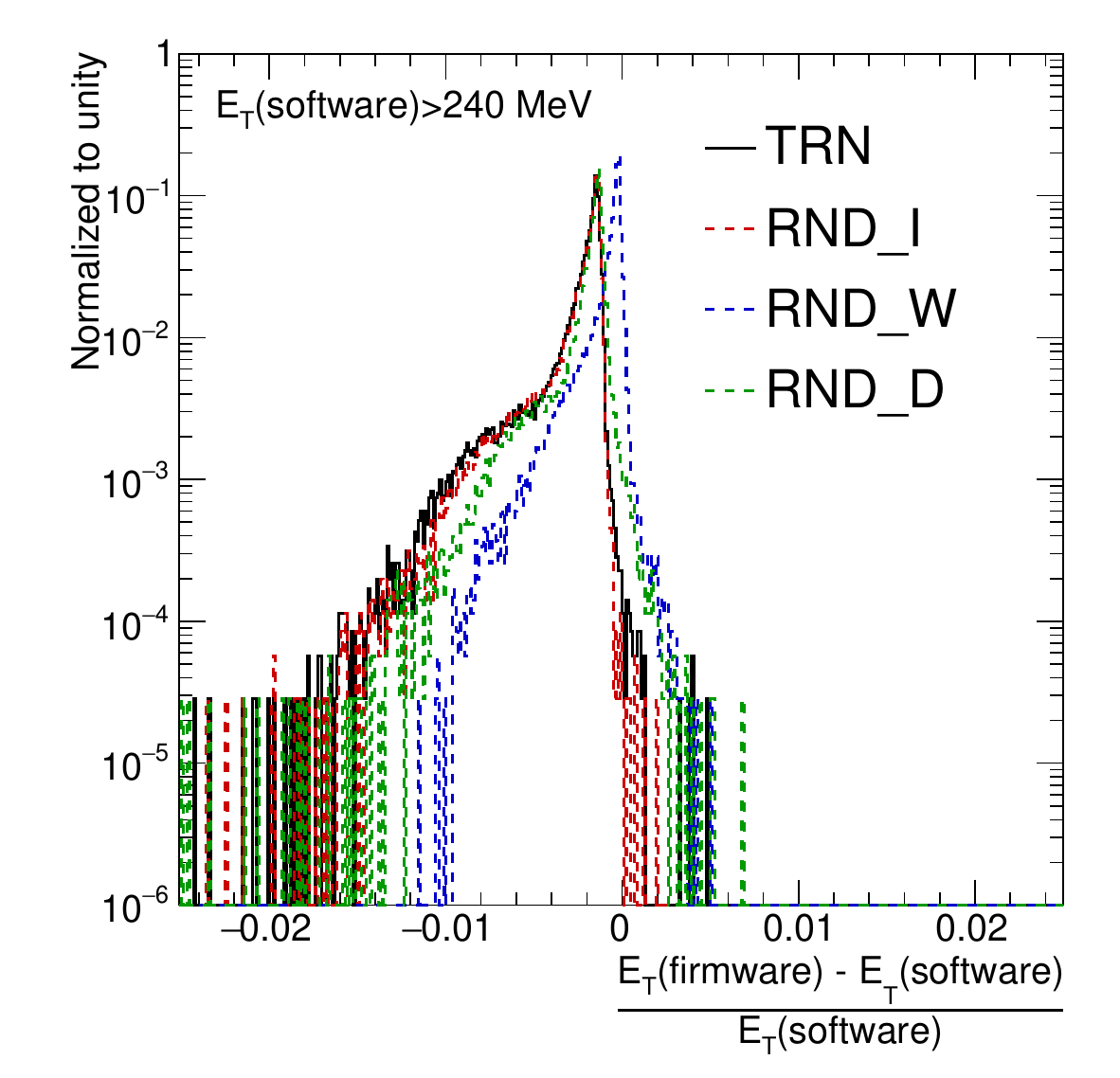}
    \includegraphics[width=0.49\textwidth]{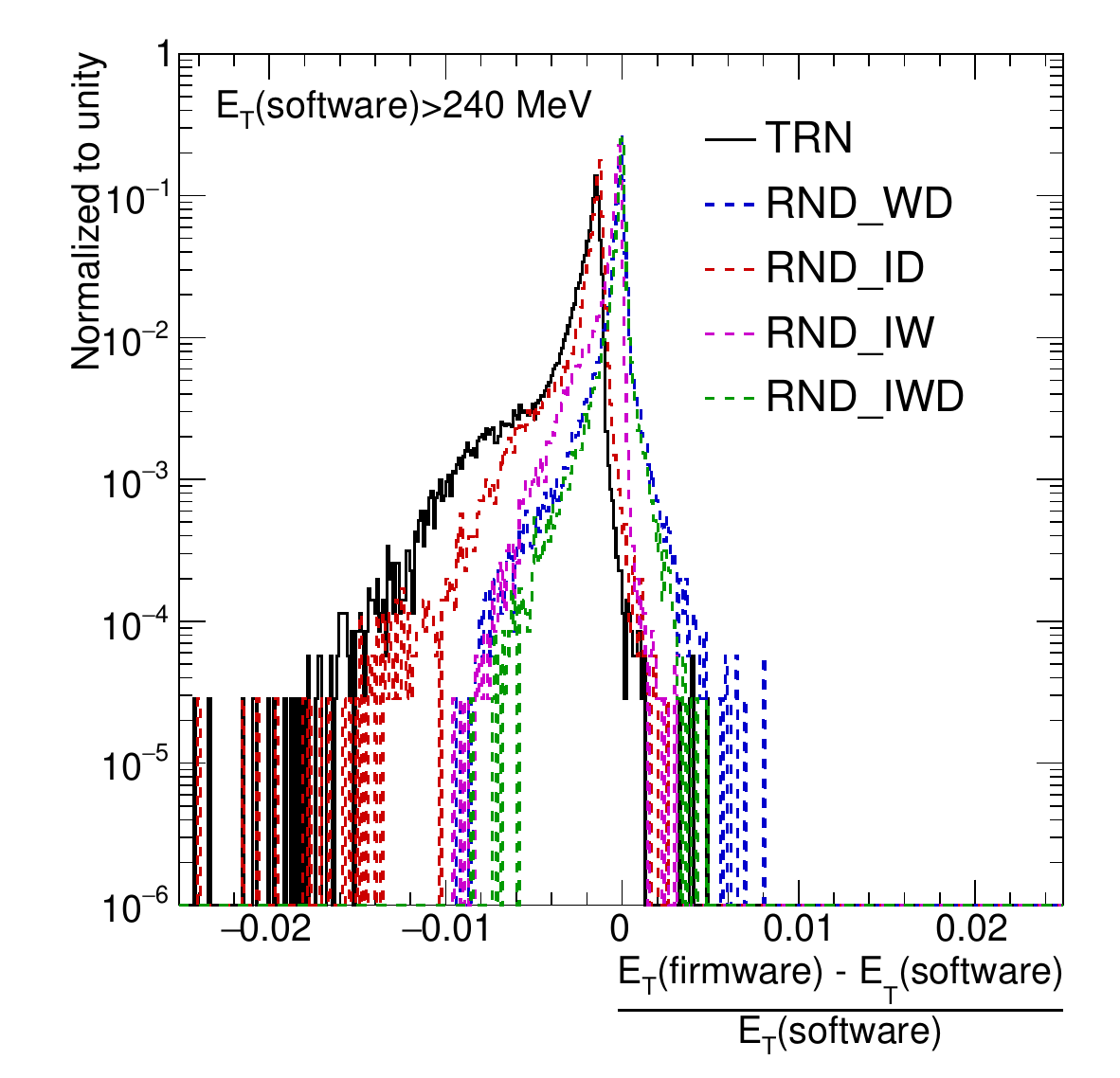}
    \caption{Resolution of the transverse energy (E$_\mathrm{T}$) computed in firmware with respect to the one computed in software. The E$_\mathrm{T}$(firmware) is computed with different implementations of the quantisation for the three data categories of the network.
    For each test, the letters I, W, and D indicate that the RND is applied for the internal data category, the weights, and the input/output data, respectively, while the TRN mode is applied by default in all other categories. A lower cut of 240 MeV is applied on $E_\mathrm{T}$(software) to remove low energies below the 3$\sigma$ noise level as described in~\cite{larnn}.}
    \label{fig:quantization_diff_types_reso}
\end{figure}

To further optimise the firmware implementation, a mix of quantisation procedures are used for different data categories. Figure~\ref{fig:quantization_diff_types} shows the resource usage and the latency while applying the TRN and RND quantisations on different data categories. The rounding of the weights does not require any additional resources in the FPGA since it can be done in software before loading these weights into the FPGA. The input data will be digitised and quantised in the frontend boards and does not require additional resources for rounding in the FPGA. The Rounding of the simulated input data is thus also done offline before loading it into the FPGA. Rounding the output data induces a slight increase in the latency. Rounding the internal data categories increases the needed resources and the latency significantly. Figure~\ref{fig:quantization_diff_types_reso} shows a comparison of the transverse energy computed in firmware and the one computed in software depending on which of the data categories is rounded. One can see that it is important to round the weights and input/output categories, while rounding the internal data category does not have any significant impact on the resolution. The root mean square (RMS) of the TRN distribution is 0.2\%. It decreases to 0.07\% if all categories are rounded (RND\_IWD). The RMS becomes slightly worse (0.09\%) if only the weights and the input/output are rounded (RND\_WD). Therefore, the TRN mode is used for the internal data computation while the RND is used for the other data categories. These optimisations allow significant improvement of the resolution of the firmware at a low resource and latency cost.

\subsection{Implementation of the neural network}
\label{sec:matrix}

 \begin{figure}[!htb]
    \centering
    \includegraphics[width=0.6\textwidth]{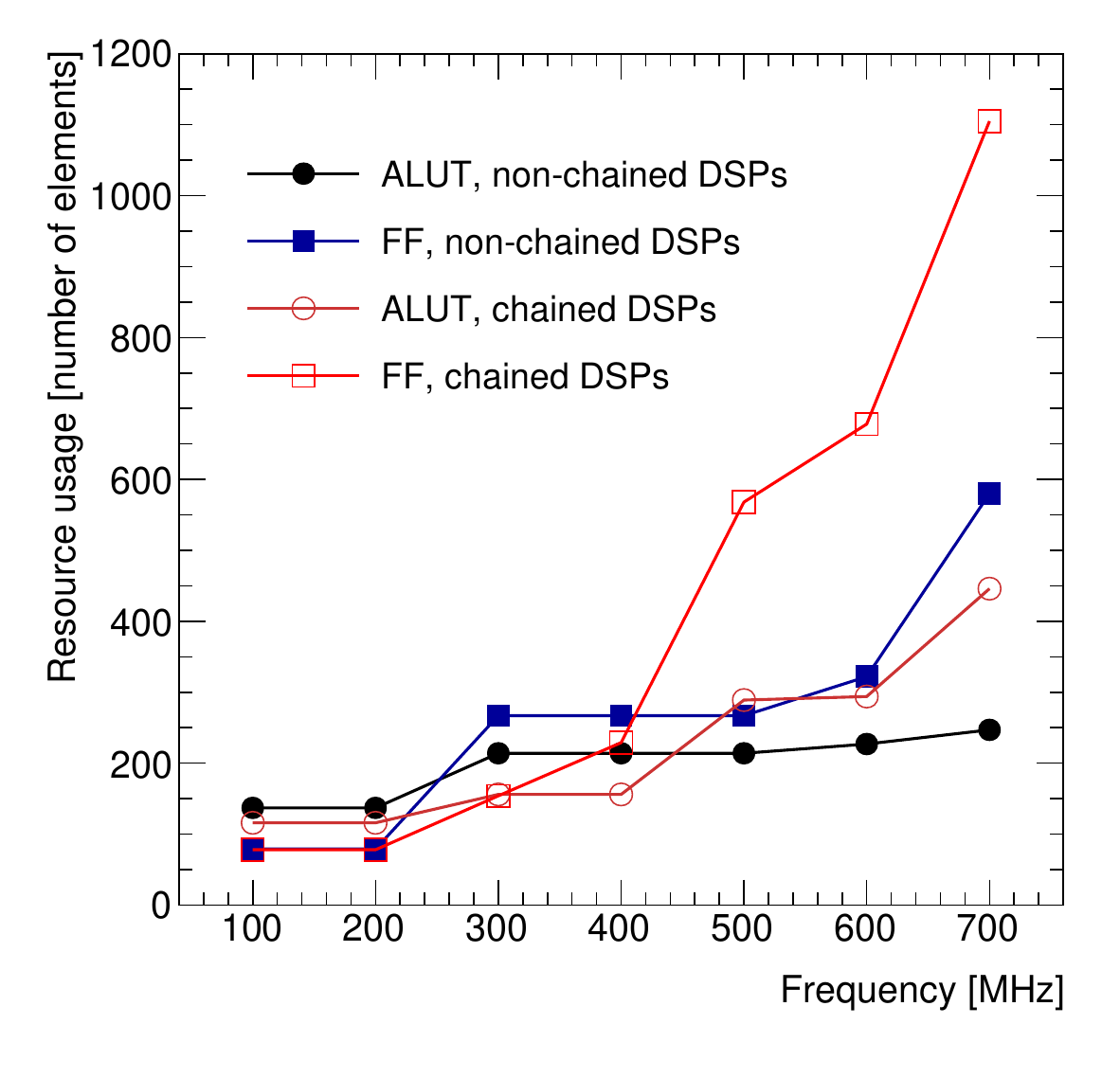}
    \caption{Resource usage for ALUTs and FFs depending on the frequency for two implementations of a matrix multiplication with and without chained DSPs. The maximum frequency of an MLAB is 450 MHz. Above this frequency, there is a significant increase of the logic resource usage for the chained DSP implementation.}
    \label{fig:chained_dsp}
\end{figure}

The computation inside the neural network requires 304 multiplications and 231 additions. The multiplications are implemented inside the DSPs, and the additions are initially implemented using ALUTs and FFs. The DSP allows summing the output of the two multiplications internally. The usage of such functionality reduces the number of additions implemented in the ALUTs and FFs from 231 to 131. Furthermore, the DSP component contains an additional adder that can take an external input. It is possible to chain two DSPs to sum their outputs by using the output of the first DSP in the additional adder of the second DSP.  By doing this, it is necessary to synchronise the DSPs so that their results arrive at the same time to the additional adder of the second DSP. Therefore, the second DSP in the chain should be shifted by 1 clock cycle. The same procedure is applied to more DSPs to build a full chain. In such case, all additions can be implemented in DSPs.

To perform the timing shift, each input of the DSPs needs an additional level of registers to delay the results. This is done by combining several ALUTs to create a memory logic array block (MLAB) to implement a first in first out (FIFO) memory. However, the MLAB frequency is limited to 450 MHz in read-during-write mode needed for FIFO implementation. At higher frequencies the synchronisation cannot be implemented in MLAB, in such case the delays are implemented in basic ALUTs and FFs which increase the number of needed logic elements significantly. At high frequency, more ALUTs and FFs are needed to synchronise the DSPs in chained mode than to implement the additions in a non-chained mode. Figure~\ref{fig:chained_dsp} shows the FPGA resource usage as function of the frequency for DSPs used in chained or non-chained mode for one matrix multiplication. The chained mode is advantageous below 450 MHz. Above this frequency, it is more advantageous to use logic elements to do the additions than to chain DSPs. This will be the option retained for the implemented RNN since we seek higher frequencies to increase the multiplexing and thus reduce the overall resource usage.

\subsection{Results of the HLS Implementation}

The designed RNN HLS code is compiled in Intel HLS and Quartus \cite{quartus}. The results of these two compilations are given in table~\ref{tab:hls_results}. For one implemented network the design can run at 455 MHz which a priory allows multiplexing up to 11 channels per network for a data input rate of 40 MHz. However, with a multiplexing of 10 the maximum frequency reached is 393 MHz while the needed frequency is 400 MHz for the firmware to run without timing violations. The maximum frequency is reduced when applying multiplexing due to the additional weights that are needed by the network to perform the computation for several channels.
Up to 37 networks can fit in the FPGA which leads to the usage of 100\% of the available DSP resources and most of the logic resources. The logic resources are given in terms of ALUTs and FFs in the HLS report and adaptive logic modules (ALM) for the Quartus report. The ALMs are the actual physical resources in the FPGA and each ALM can be configured to be used as two ALUTs or 4 FFs.
The HLS design does not allow reaching the required 384 channels processed in one FPGA, even with full utilisation of resources in the FPGA. In practice only part of these resources will be available for the transverse energy computation as discussed in section~\ref{sec:results_vhdl}. Additional optimisations are needed to fulfil the requirements of the LAr calorimeter. Furthermore, the latency of this HLS firmware is 277.5 ns which is significantly larger than the required 125 ns. The HLS design was optimised for the maximum possible frequency which leads to additional registers added by the compiler to meet the required timing constraints. This in turn increases the latency of the design.

\begin{table}[!htb]
    \centering 
    \caption{Resource usage and the maximum frequency (Fmax) given by the Intel HLS and Quartus compiler reports for one and 37 implemented networks in the FPGA. The multiplexing is set to 10 so that each network instance can process 10 independent channels.}
    \resizebox{\textwidth}{!}{
        \begin{tabular}{c|c|c|c|c|c|c|c|c|c}
          & Networks & Multiplexing & Channels & ALUTs & FFs & ALMs & DSPs & Memory & Fmax\\
        \hline
        & 1 & 10 & 10 & 0,6{\%} & 0.7\% & - &  3{\%} & 0,3{\%} &  - \\
        HLS & 37 & 10 & 370 & 22.9\% & 25.9\% & - &  100{\%} & 11{\%} &  - \\
        \hline
        & 1 & 10 &  10 & - & - & 2,4{\%} & 3{\%} & 0,3{\%} & 455 MHz \\
        Quartus & 37 & 10 &  370 & - & - & 90{\%} & 100{\%} & 11{\%} & 393 MHz \\
        \hline
        \end{tabular}
    }
    \label{tab:hls_results}
\end{table}

\section{VHDL Implementation}
 The HLS implementation adds an additional level of abstraction that allows fast and efficient optimisation of the network parameters and firmware implementation. However, this additional level of abstraction prevents some finer optimisations that are possible in VHDL. These finer optimisations allow to meet the specifications, which is not possible for the HLS implementation. That is why VHDL is used for the final optimisation and placement of the RNN firmware implementation.
 
\subsection{Reuse of common computations between RNN cells}
\label{sec:firstcell}
As shown in figure \ref{fig:RNN}, the Kernel Weight Multiplication and the Bias Addition depend only on the input of the neural network and do not depend on the past network state. However, these two computations done in the second cell at a given time $t$ are identical to the computations in the first cell at time $t-1$. More generally, these computations in cell $n$ at time $t$ are identical to the computations in the first cell at time $t-n-1$. One can do these two computations one time at the first cell and propagate the output to the other four cells at the proper time. Moreover, the first cell does not need the Recurrent Weight Matrix Multiplication since there is no previous RNN state at this level. Therefore, this computation is removed from the first cell. These two optimisations allow a reduction in the number of used DSPs by 10\%. The ALM usage is also reduced by 21\% due to the removal of the duplicated Bias Additions.
   
\subsection{Placement constraints}
\label{sec:placement}

Several instances of the neural network are needed to process all 384 channels. In a given compilation each instance have a different placement shape. Moreover, these shapes change between compilations due to the randomisation in Quartus. This complicates the optimisation of the timing critical paths that is needed to reach higher frequencies and thus higher multiplexing.

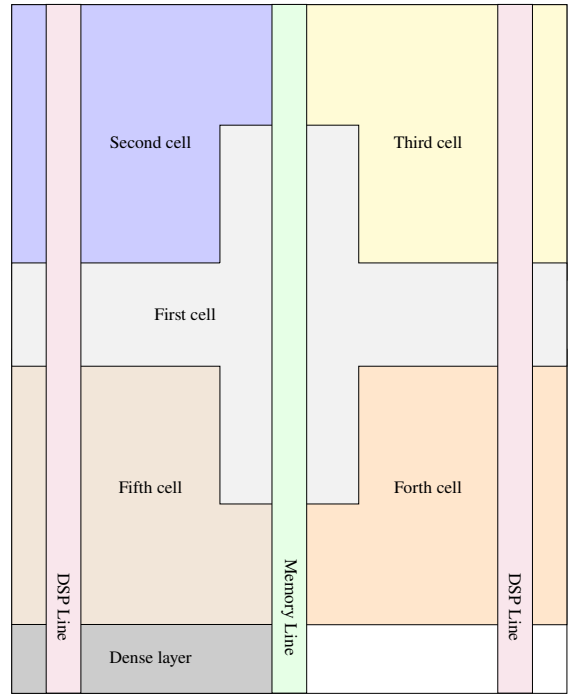
\begin{figure}[!htb]
    \centering
    \resizebox{0.5\textwidth}{!}{
        \input{fig10.tex}
    }
    \caption{Schematic view of the optimised placement of the Vanilla RNN showing the 5 cells and the dense layer with respect to the memory and DSP lines inside the FPGA.
    }
    \label{fig:Placement}
\end{figure}

Placement constraints are used to force the same placement shape of the implemented neural network. Thus, all instances of the neural network have the same shape which simplifies the optimisation of the critical paths. Moreover, the placement of the 5 network cells is optimised to minimise the distance between connected cells. The shape of the neural network is shown in figure~\ref{fig:Placement}. The first cell is placed in the middle since it is connected to all other four cells as described in section~\ref{sec:firstcell}. The other cells are placed around the first cell and ordered to reduce the distance between consecutive cells. The dense layer is placed next to the fifth cell.

All cells use the same set of recurrent weights except for the first cell that does not have a recurrent block as explained in section~\ref{sec:firstcell}. These weights are stored in memory (M20K blocks) and are directly connected to the DSPs to make the matrix multiplications. To reduce the mean distance between the M20k block and each DSP the recurrent weights are duplicated. Each set of weights is used for two cells instead of four.

These placement constraints allow increasing the maximum frequency from 434 MHz to 492 MHz with 28 instances of the RNN implemented in the FPGA.

\subsection{Incremental compilation}
Quartus compilation is composed of three parts: Analysis and Synthesis, Fitting, and Timing Analysis.
The Analysis and Synthesis will translate the VHDL code into RTL. The RTL code creates a high-level representation of a circuit. The Fitter will place and route the design into the FPGA. It will determine the required resources and the wiring between the different components. The Timing Analysis will determine the maximum frequency that the FPGA can reach with a given design.

Quartus can divide a firmware design into multiple partitions.
It also provides the possibility to preserve the results of a given compilation at different steps of the process for each partition. To further increase the maximum frequency the firmware design is partitioned in a way that each partition corresponds to one neural network instance.
A sequence of compilations are performed and each partition that reaches the target frequency is preserved while the others are recompiled.
Several combinations of target frequencies and numbers of neural networks instances are tested. We converged on a configuration with a target frequency of 560 MHz and 28 instances of the RNN. This allows to run the RNN with a multiplexing of 14 and reach the required number of channels to be processed in one FPGA.
Four compilations are needed to reach the target frequency in this configuration, while 18 out of the 28 instances reach the target frequency at the first compilation. 

\subsection{Results of the VHDL implementation}
\label{sec:results_vhdl} 
The final firmware implemented in VHDL contains 28 instances of the vanilla RNN each with a multiplexing of 14 which allows covering 392 channels. The results from this firmware are summarised in table~\ref{tab:vhdl_results}. This firmware can run at 561 MHz while the required frequency for a multiplexing of 14 is 560 MHz. The final requirements on the resource usage are not available since the full final phase-II firmware is not yet available. However, we require a resource usage of the neural network block similar to the transverse energy computation block used for the phase-I upgrade \cite{phase1_tdr} of the LAr calorimeter which is currently in operation. This block uses about 70\% of the DSPs and 30\% of the logic. The optimised RNN resource usage is within these requirements as shown in table~\ref{tab:vhdl_results}. The latency of the firmware is 65 clock cycles corresponding to 116 ns which is within the 125 ns required latency. The optimisations done in VHDL do not affect the results of the energy computation; therefore, the firmware resolution stays below 0.1 \% as it was for the HLS implementation.
  
\begin{table}[!htb]
    \centering
    \caption{Resource usage and the maximum frequency (Fmax) given by the Quartus compiler report for 28 implemented networks in the FPGA. The multiplexing is set to 14 so that each network can process 14 independent channels.}
    \begin{tabular}{c|c|c|c|c|c|c|c}
    \hline
    Type & Networks & Multiplexing & Channels & ALM & DSP & Memory & Fmax\\
    \hline
    Specification & - & - & 384 & 30{\%} & 70{\%} & 30{\%} & 560 MHz\\
    \hline
    Quartus & 28 & 14 & 392 & 18{\%} & 66{\%} & 16{\%} & 561 MHz \\
    \hline
    \end{tabular}
    \label{tab:vhdl_results}
\end{table} 
    
\section{Conclusion}

The phase-II upgrade of the LAr calorimeter allows the unique opportunity to implement neural networks in FPGAs in order to improve the computation of the transverse energy deposited in the calorimeter.
This report presents the implementation of a vanilla RNN in Stratix 10 FPGAs from Intel. The firmware is first developed in HLS for fast prototyping and optimisation of the network architecture. The implemented network in firmware matches the computation of the deposited transverse energy in software with a resolution better than 1\textperthousand~ after the optimisation of the bit width of the fixed point representation and the quantisation of the arithmetic operations. However, the HLS implementation could neither reach the required frequency to allow the implementation of 384 channels per FPGA nor the required latency. VHDL is used to further optimise the HLS implementation and to add constraints on the placement allowing to reach the required specifications of the firmware. The final result is a firmware containing 28 instances of the RNN capable of running at 560 MHz with a multiplexing of 14 and a latency of 65 clock cycles (116 ns). This firmware requires less than 70\% of the available DSPs and less than 20\% of the available logic elements in the Stratix 10 FPGA.

\section{Acknowledgements}
The project leading to this publication has received funding from Excellence Initiative of Aix-Marseille Universit\'e - A*MIDEX, a French ``Investissements d’Avenir'' programme, AMX-18-INT-006 and from the French ``Agence National de la Recherche'', ANR-20-CE31-0013. This work received support from the French government under the France 2030 investment plan, as part of the Excellence Initiative of Aix-Marseille University - A*MIDEX (AMX-19-IET-008 - IPhU). This work would not be possible without the support of the Institut f{\"u}r Kern- und Teilchenphysik of the Technische Universit{\"a}t Dresden.

\end{document}

%% file: fig3.tex
\begin{tikzpicture}

    \node[draw, minimum width = 18cm, minimum height = 8cm, align=center] (RNN) at (0,0) {};
    \draw (RNN.south east) node [above left, align =center]{Vanilla RNN \\  5 cells};
    
    \draw (RNN.north west) ++ (1,-3) node(Position_first_cell){} ++ (0.25,0.25) node (Position_other_first_cell){};
    \draw (RNN.north east) ++ (0,-3) node(Position_Output){};
    \draw (Position_Output) node[right, align = center] () {Energy \\ reconstructed};
    \node[draw, fill=white, minimum width = 2cm, minimum height = 2cm, align =center, right of = Position_first_cell, node distance = 0cm, anchor = west] (First_cell){First \\ cell};
    \node[draw, fill=white, minimum width = 2cm, minimum height = 2cm, align =center, right of = First_cell, node distance = 1.75cm, anchor = west] (Second_cell){Second \\ cell};
    \node[draw, fill=white, minimum width = 2cm, minimum height = 2cm, align =center, right of = Second_cell, node distance = 1.75cm, anchor = west] (Third_cell){Third \\ cell};
    \node[draw, fill=white, minimum width = 2cm, minimum height = 2cm, align =center, right of = Third_cell, node distance = 1.75cm, anchor = west] (Forth_cell){Forth \\ cell };
    \node[draw, fill=white, minimum width = 2cm, minimum height = 2cm, align =center, right of = Forth_cell, node distance = 1.75cm, anchor = west] (Fifth_cell){Fifth \\ cell };
    \node[draw, fill=white, minimum width = 2cm, minimum height = 2cm, align =center, right of = Fifth_cell, node distance = 1.75cm, anchor = west] (Dense_layer){Dense \\ layer};
    \draw (RNN.south) node[below, align =center] (Input){Signal from one \\ LAr channel};
    \draw (RNN.south) ++ (0,2) node(Position_Input){};

    \path (First_cell.east) -- node (text_1) {/} (Second_cell.west);
    \node [above of = text_1, node distance = 0.3cm] (8_1){8};
    \draw[->] (First_cell.east) --(Second_cell.west);

    \path (Second_cell.east) -- node (text_2) {/} (Third_cell.west);
    \node [above of = text_2, node distance = 0.3cm] (8_2){8};
    \draw[->] (Second_cell.east) -- (Third_cell.west);

    \path (Third_cell.east) -- node (text_3) {/} (Forth_cell.west);
    \node [above of = text_3, node distance = 0.3cm] (8_3){8};
    \draw[->] (Third_cell.east) -- (Forth_cell.west);

    \path (Forth_cell.east) -- node (text_4) {/} (Fifth_cell.west);
    \node [above of = text_4, node distance = 0.3cm] (8_4){8};
    \draw[->] (Forth_cell.east) -- (Fifth_cell.west);

    \path (Fifth_cell.east) -- node (text_5) {/} (Dense_layer.west);
    \node [above of = text_5, node distance = 0.3cm] (8_5){8};
    \draw[->] (Fifth_cell.east) -- (Dense_layer.west);

    \path (Dense_layer.east) -- node (text_6) {/} (Position_Output.west);
    \node [above of = text_6, node distance = 0.3cm] (1_1){1};
    \draw[->] (Dense_layer.east) -- (Position_Output.center);
    \draw (RNN.south) -- (Position_Input.center);
    \draw[->] (Position_Input.center) -| (First_cell.south);
    \draw[->] (Position_Input.center) -| (Second_cell.south);
    \draw[->] (Position_Input.center) -| (Third_cell.south);
    \draw[->] (Position_Input.center) -| (Forth_cell.south);
    \draw[->] (Position_Input.center) -| (Fifth_cell.south);

\end{tikzpicture}

%% file: fig10.tex
\begin{tikzpicture}

    \node[draw, minimum width = 16cm, minimum height = 20cm, align=center] (RNN) at (0,0) {};
    \draw (RNN.north west) node(Position_second_cell){};
    \draw (RNN.north east) node(Position_third_cell){};
    \draw (RNN.east) node(Position_forth_cell){};
    \draw (RNN.west) node(Position_fifth_cell){};
    \draw (RNN.south west) node(Position_dense){};
    \node[draw, minimum width = 8 cm, minimum height = 8cm, below of = Position_second_cell, anchor = north west, node distance = 0 cm, fill=blue!20] (Second_cell){\Large Second cell};
    \node[draw, minimum width = 8 cm, minimum height = 8cm, below of = Position_third_cell, anchor = north east, node distance = 0 cm, fill=yellow!20] (Third_cell){\Large Third cell};
    \node[draw, minimum width = 8 cm, minimum height = 8cm,  below of = Position_forth_cell, anchor = north east, node distance = 0 cm, fill=orange!20] (Forth){\Large Forth cell};
    \node[draw, minimum width = 8 cm, minimum height = 8cm, below of = Position_fifth_cell, anchor = north west, align = center, node distance = 0 cm, fill=brown!20] (Fifth_cell){\Large Fifth cell};
    \node[draw, minimum width = 8 cm, minimum height = 2cm, below of = Position_dense, anchor = south west, node distance = 0 cm, fill=black!20, align =center] (Dense_layer){\Large Dense layer};
    \draw [fill = black!5] (RNN.center) + (-3,1) node(){\Large First cell} ++ (-2,2.5)  |- ++ (4,4) |- ++ (6,-4) |- ++ (-6,-3) |- ++ (-4,-4) |- ++ (-6,4) |- ++ (6,3);
    \draw (RNN.north west) ++ (1.5,0) node(Position_DSP1){};
    \node[draw, minimum width = 1cm, minimum height = 20cm, align=center, text height = 15cm, anchor=north, below of = Position_DSP1, node distance = 10cm, fill=purple!10] (DSP1){\rotatebox{-90}{\Large DSP Line}};
    \draw (RNN.north east) ++ (-1.5,0) node(Position_DSP2){};
    \node[draw, minimum width = 1cm, minimum height = 20cm, align=center, text height = 15cm, anchor=north, below of = Position_DSP2, node distance = 10cm, fill=purple!10] (DSP2){\rotatebox{-90}{\Large DSP Line}};
    \draw (RNN.north) node(Position_RAM){};
    \node[draw, minimum width = 1cm, minimum height = 20cm, text height = 15cm, align=center, anchor=north, below of = Position_RAM, node distance = 10cm, fill=green!10] (RAM){\rotatebox{-90}{\Large Memory Line}};

\end{tikzpicture}